\documentstyle{mn}
\newif\ifAMStwofonts
\ifoldfss
  \ifCUPmtlplainloaded \else
    \NewTextAlphabet{textbfit} {cmbxti10} {}
    \NewTextAlphabet{textbfss} {cmssbx10} {}
    \NewMathAlphabet{mathbfit} {cmbxti10} {} % for math mode
    \NewMathAlphabet{mathbfss} {cmssbx10} {} %  "   "    "
  \fi
  \ifAMStwofonts
    \ifCUPmtlplainloaded \else
      \NewSymbolFont{upmath} {eurm10}
      \NewSymbolFont{AMSa} {msam10}
      \NewMathSymbol{\upi}     {0}{upmath}{19}
      \NewMathSymbol{\umu}     {0}{upmath}{16}
      \NewMathSymbol{\upartial}{0}{upmath}{40}
      \NewMathSymbol{\leqslant}{3}{AMSa}{36}
      \NewMathSymbol{\geqslant}{3}{AMSa}{3E}

    \fi
  \fi
\fi % End of OFSS

\ifnfssone
  \newmathalphabet{\mathit}
  \addtoversion{normal}{\mathit}{cmr}{m}{it}
  \addtoversion{bold}{\mathit}{cmr}{bx}{it}
  \newmathalphabet{\mathbfit} % math mode version of \textbfit{..}
  \addtoversion{normal}{\mathbfit}{cmr}{bx}{it}
  \addtoversion{bold}{\mathbfit}{cmr}{bx}{it}
  \newmathalphabet{\mathbfss} % math mode version of \textbfss{..}
  \addtoversion{normal}{\mathbfss}{cmss}{bx}{n}
  \addtoversion{bold}{\mathbfss}{cmss}{bx}{n}
  \ifAMStwofonts
    \ifCUPmtlplainloaded \else
      \UseAMStwoboldmath
      \makeatletter
      \new@mathgroup\upmath@group
      \define@mathgroup\mv@normal\upmath@group{eur}{m}{n}
      \define@mathgroup\mv@bold\upmath@group{eur}{b}{n}
      \edef\UPM{\hexnumber\upmath@group}
      \new@mathgroup\amsa@group
      \define@mathgroup\mv@normal\amsa@group{msa}{m}{n}
      \define@mathgroup\mv@bold\amsa@group{msa}{m}{n}
      \edef\AMSa{\hexnumber\amsa@group}
      \makeatother
      \mathchardef\upi="0\UPM19
      \mathchardef\umu="0\UPM16
      \mathchardef\upartial="0\UPM40
      \mathchardef\leqslant="3\AMSa36
      \mathchardef\geqslant="3\AMSa3E
    \fi
  \fi
\fi % End of NFSS release 1

\ifnfsstwo
  \DeclareMathAlphabet{\mathbfit}{OT1}{cmr}{bx}{it}
  \SetMathAlphabet\mathbfit{bold}{OT1}{cmr}{bx}{it}
  \DeclareMathAlphabet{\mathbfss}{OT1}{cmss}{bx}{n}
  \SetMathAlphabet\mathbfss{bold}{OT1}{cmss}{bx}{n}
  \ifAMStwofonts
    \ifCUPmtlplainloaded \else
      \DeclareSymbolFont{UPM}{U}{eur}{m}{n}
      \SetSymbolFont{UPM}{bold}{U}{eur}{b}{n}
      \DeclareSymbolFont{AMSa}{U}{msa}{m}{n}
      \DeclareMathSymbol{\upi}{0}{UPM}{"19}
      \DeclareMathSymbol{\umu}{0}{UPM}{"16}
      \DeclareMathSymbol{\upartial}{0}{UPM}{"40}
      \DeclareMathSymbol{\leqslant}{3}{AMSa}{"36}
      \DeclareMathSymbol{\geqslant}{3}{AMSa}{"3E}
    \fi
  \fi
\fi % End of NFSS release 2

\ifCUPmtlplainloaded \else
  \ifAMStwofonts \else % If no AMS fonts
    \def\upi{\pi}
    \def\umu{\mu}
    \def\upartial{\partial}
  \fi
\fi

\title{Sub-mm and near-IR observations of galaxies selected at 170\,$\mu$m.}
\author[Sajina et al.]
       {A.~Sajina$^1$, 
  C.~Borys$^2$, S.~Chapman$^2$, H.~Dole$^3$, M.~Halpern$^1$, 
  G.~Lagache$^4$, 
\newauthor 
J.-L.~Puget$^4$, D.~Scott$^1$\\
  $^1$Department of Physics \& Astronomy, University of British Columbia, 6224 Agricultural Road, Vancouver, BC V6T1Z1, Canada\\
  $^2$Department of Physics, California Institute of Technology, Pasadena, CA 91125, USA\\
  $^3$Steward Observatory, University of Arizona, 933 N Cherry Ave, Tucson, AZ 85721, USA\\
  $^4$Institut d'Astrophysique Spatiale, Universite Paris Sud, Bat. 121,  91405 Orsay Cedex, France\\
  	}

\date{Draft version 17 March 2003}

\pagerange{\pageref{firstpage}--\pageref{lastpage}}
\begin{document}

\maketitle

\label{firstpage}

\begin{abstract}
We present results from JCMT sub-mm observations of sources selected from the
{\sl ISO\/} FIRBACK (Far-IR BACKground) survey, along with UKIRT near-IR
imaging of a sub-sample.  This gives valuable insight into the brightest
$\sim$10\% of galaxies which contribute to the Cosmic Infrared Background (CIB).
We estimate the photometric redshifts and
luminosities of these sources by fitting their Spectral Energy Distributions
(SEDs).  The data appear to show a bimodal galaxy distribution, with
normal star-forming galaxies at $z\simeq0$, and a much more luminous population at $z\sim0.4$--0.9.  These are similar to the ultraluminous
infrared galaxies which are found to evolve rapidly with redshift in other
surveys.  The detectability threshold of FIRBACK biases the sample away from much higher redshift ($z\stackrel{>}{_{\sim}}1.5$) objects.  Nevertheless, the handful of
$z\sim0.5$ sources which we identify are likely to be the low-$z$ counterparts
of the typically higher-$z$ sources found in blank field sub-mm observations.
This sub-sample, being much more nearby than the average SCUBA galaxies, has the virtue of being relatively easy to
study in the optical.  Hence their detailed investigation could help
elucidate the nature of the sub-mm bright galaxies.
\end{abstract}

\begin{keywords}
infrared: galaxies -- submillimetre.
\end{keywords}

\section{Introduction}

What makes up the Cosmic Infrared Background (CIB) detected from the
{\sl COBE\/}-FIRAS data \cite{p96,f98,l99,l00,h98,dw98,fk00}?
This remains an open question, and details of galaxy types, their redshift
distribution, and how they appear in other wavebands, remain sketchy.
The FIRBACK (Far-IR BACKground) survey \cite{p99,d01} addressed this question
by performing some of the deepest blank field {\sl ISO\/}
surveys at 170\,$\mu$m, near the peak of that radiation.
About 200 sources were detected above 3\,$\sigma$ (=135\,mJy)
accounting for about 7$\%$ of the nominal background value.

In general far-IR sources such as the FIRBACK ones sample the low-to-moderate
redshift regime, and thus provide a link between the local Universe and
high-$z$ sources, such as the SCUBA-bright `blank-sky' population (see Blain et al.~2002 for a review).
Understanding the nature of these sources, their emission mechanisms, and
their dust properties is crucial to our understanding of galaxy formation
and evolution from high redshift until today.  This in turn informs us about
the cosmic background, as well as the nuclear activities, star formation
distributions, and the role of dust obscuration in star formation through a
large fraction of the history of the Universe. \\
In order to better understand the sources detected by FIRBACK we have been
carrying out follow-up observations with SCUBA at 450\,$\mu$m and 850\,$\mu$m.
Detection by {\sl ISO\/} at 170\,$\mu$m ($S_{\rm{170}}$$>$135\,mJy),
means that a strong bias away from high-$z$ objects is present,
although we still expect to detect objects out to $z\simeq1$.
The FIRBACK galaxies represent the brightest contributors to the CIB, and
are a different selection than typical SCUBA galaxies.  `Blank-sky'
sub-mm bright galaxies, although so far accounting for up to
50\% of the {\it sub-mm\/} background (e.g.~Cowie, Barger, \& Kneib~2002),
make up an insignificant fraction of the total CIB.  The combination of
far-IR and sub-mm observations is thus very powerful in establishing a link
between high-$z$ dusty starbursts and their local counterparts.

Observationally, without redshifts, it has been difficult to distinguish between cooler local starbursts and warmer, more luminous sources at higher $z$.  This is because the spectral energy distributions (SEDs), for a fixed emissivity index $\beta$, are degenerate in the parameter combination $(1+z)/T_{\rm{d}}$. There are additional complications of course caused by variations in $\beta$, $T_{\rm{d}}$, and luminosity, changing the shape of the SED.  Fundamentally, our understanding of dust in extragalactic sources, its properties, and interaction with the radiation field is poor, which is a major impediment in our interpretation of the observational evidence.

The only way to approach these issues is by detailed multiwavelength studies of samples representing key elements in the above puzzle.
 
The data we present in this paper constitute a far-IR selected sample. This sample is now large enough, and with wide enough wavelength coverage, including near-IR, far-IR, sub-mm and radio observations, to be able to tackle some of these issues.  We do this through a combination of direct SED fitting, statistical analysis, consistency with other observations and comparison with model predictions, trying to use the minumum number of a priori assumptions.  In essence, we start with the assumption that our sources represent a uniform population, which we demonstrate are most likely local starforming galaxies. This model breaks down for a handful of our sources, which is probably due to their being at somewhat higher redshifts (and more luminous) than the rest of the sample. Since our sample is representative of FIRBACK as a whole, we have thus improved our knowledge of the nature of the brightest sources making up the CIB.

We focus largely on the conclusions deriving from the sub-mm observations. Discussion of the individual properties of FIRBACK galaxies in general, including issues relating to the identification of the sources will be discussed elsewhere \cite{lg03}. 

Throughout the paper we assume a flat Universe with $H_{0}$=75~km~s$^{-1}$~Mpc$^{-1}$, $\Omega_{\rm{M}}$=0.3, and $\Omega_{\Lambda}$=0.7. 
\section[]{The Data}
\subsection{Sample}
Targets were selected from the FIRBACK ($>3\sigma$) catalogue (Dole et al. 2001) in the ELAIS N1 field. Confusion remains a major issue with a beam size $\sim$\,90\,arcsec. In our selection we aimed at a uniform sample that would be as representative of the FIRBACK population as possible. Our main selection criterion has been the availability of radio detections \cite{ci99} inside the beam (required for JCMT pointing). 

We exclude FIRBACK sources with radio-bright counterparts which are rare and not representative of the typical FIRBACK source, as well as sources with several radio detections per beam \footnote{This criterion was not applied too strictly and a few sources (N1-008, N1-029, N1-041) do in fact have two radio sources inside the beam.}. This left us with 41 possible sources to draw from -- we have followed up 30 (and additionally one N2 source) -- listed in Table\,1. 
Due to this broadness of selection criteria we believe our sample represents a fair cross-section of the FIRBACK population, rather than focusing on a specific sub-population. This can be seen clearly in Fig.~1, where there is close agreement between the distribution of $S_{170}/S_{1.4\rm{GHz}}$ for all 41 possible sources with that of the 30 sources in our sample.  

\begin{figure}
\centering
\vspace*{8cm}
\leavevmode
\includegraphics{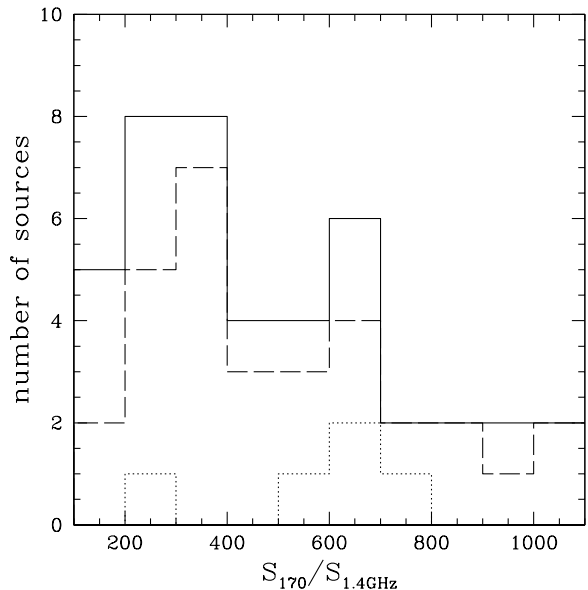}
\caption{Distribution of the $S_{170}/S_{1.4\rm{GHz}}$ ratio. The solid histogram represents the 41 sources from the FIRBACK catalogue which meet our selection criteria. The dashed histogram is our sub-sample of 31 targets. The dotted line shows our high-$z$ candidates (discussed in Section 4).}
\end{figure}

\subsection{SCUBA sub-mm observations}
The observations presented here were taken with the Sub-mm Common User Bolometer Array -- SCUBA \cite{h99} instrument on the James Clerk Maxwell Telescope (JCMT) in March 1999 and in March and May 2001. In order to avoid biasing our data, we attempted to observe each source until a predetermined rms ($\sim\,1.5\,$mJy) was reached, irrespective of whether the source appeared to be a possible detection or not. The March 2001 data were taken in exceptional grade 1 weather ($\tau_{225}\sim0.04$, and as low as 0.02), whereas the 1999 data were taken in merely `good' weather  ($\tau_{225}\sim0.07$). Throughout our observations we used the 2-bolometer chopping mode. This involves chopping in array coordinates in order to always align one negative beam exactly with a specific off-centre bolometer with the other being partially aligned. Thus the negative beams can be folded in, improving the rms by a factor of up to $\sqrt{2/3}$. 
 Unfortunately, for our 2001 run, software problems with the new telescope control system resulted in SCUBA not properly chopping onto another bolometer, making the negative beams unrecoverable. Thus the 2001 data presented here are from the central bolometer only, whereas the 1999 data have the negative beams folded in. \\
The data were reduced using the {\sc SURF} package \cite{j98}, and also with custom-written code. The extinction correction was performed using sky-dip observations whenever available, and with a derived optical depth from the $\tau_{\rm{\sc cso}}-\tau_{\rm{\sc scuba}}$ relations \cite{a00} otherwise. The sky was subtracted as a mean level for the entire array at each jiggle, excluding the signal bolometers as well as any bad bolometers, as revealed by excessive noise levels. A calibration uncertainty of a maximum of 20$\%$ exists (based on differences among the nightly calibration values), however it has little effect on the signal-to-noise ratios, especially for the 2001 data where each source was typically observed in a single night. The pointing uncertainty was $\sim$\,2\,arcsec. The SCUBA 850\,$\mu$m beam FWHM is $\sim$\,15\,arcsec, whereas the 450\,$\mu$m beam is $\sim$\,8\,arcsec.\\
The derived 450\,$\mu$m and 850\,$\mu$m flux densities and associated 1\,$\sigma$ errors are listed in Table~1. 
\subsection{UKIRT near-IR observations}
Most of our SCUBA detected FIRBACK sources could not be identified in
DPOSS images. As the sources are expected to be extinguished by dust
and therefore have red spectra, we obtained deep
observations in the $K$-band at UKIRT, using the Fast Track Imager (UFTI)
for maximum sensitivity to obscured components.
The small UFTI field (50\,arcsec$\times$50\,arcsec) was centred on
the source positions taken from the radio/SCUBA identifications.
Each source was imaged for a total of 1800\,s, with individual exposures of 60\,s each, reaching a limiting magnitude in a 2\,arcsec diameter aperture of $K=20.4$\,(5$\sigma$). The fast tip/tilt, adaptively corrected imaging resulted in seeing better than median conditions, at $\simeq$0.4\,arcsec FWHM.  Data were reduced using the Starlink, UKIRT/UFTI image processing tools under the {\sc oracdr} environment \cite{br00}. We wrote custom {\sc oracdr} scripts to optimize point source sensitivity in our essentially blank field observations, creating flat fields from each 9-point dither, and high signal-to-noise thermal background images from 60 minutes of data centered around the observing period of each target. 
\begin{figure*}
\centering
\vspace*{9.5cm}
\leavevmode
\includegraphics{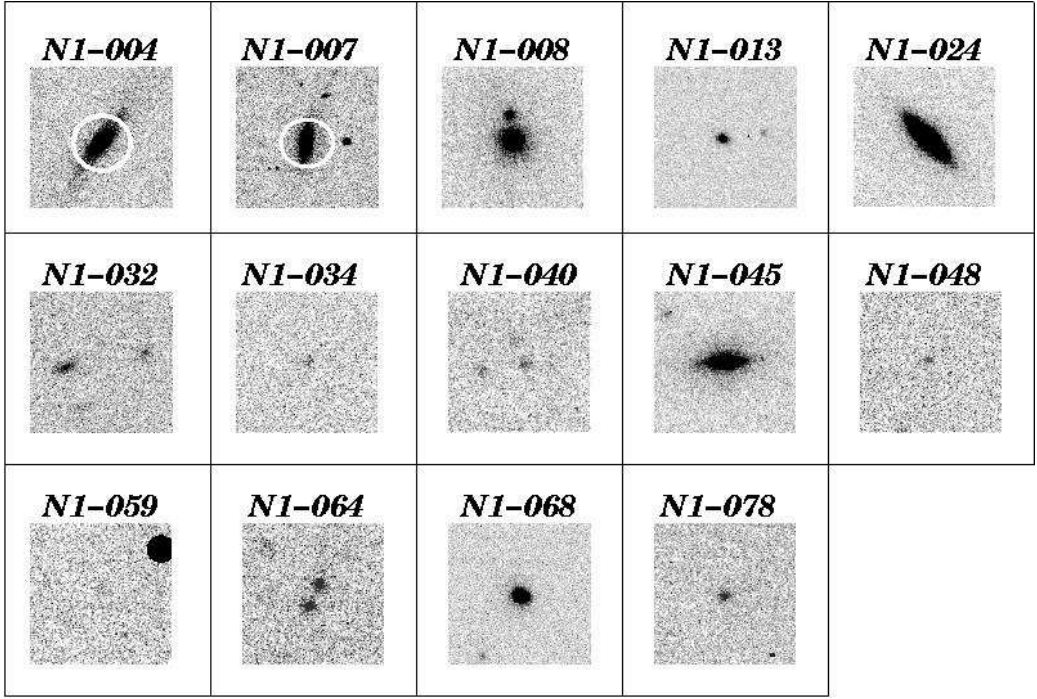}
\caption{\footnotesize
The available UKIRT $K$-band images. The images are centered on the radio position, and are about 15$\times$15\,arcsec or roughly a SCUBA 850\,$\mu$m beam size. Exceptions are N1-004 and N1-007, where the white circles have a diameter of $\sim$15\,arcsec.
}
\end{figure*}
 Fig.~2 shows the available UKIRT images. For the rest of our sample we use data from the 2MASS catalogue (for sources with $K\sim$\,14--15 we estimated their magnitudes directly from the calibrated catalogue images via aperture photometry in {\sc gaia}). 
\subsection{Far-IR and radio fluxes}
{\sl IRAS} 100\,$\mu$m fluxes were obtained using the {\sc xscanpi} facility. We quote them here for the sake of completeness, as well as to help in comparison with local {\sl IRAS} galaxies (which are typically $>$1\,Jy).  The errors quoted are purely statistical; however, at this faint level systematic errors (including, among other things IR cirrus and mapping artefacts) dominate the flux estimates, to the point of making them of little statistical use. ISOPHOT 90\,$\mu$m detections also exist for a few of the objects, as well as mid-IR ISOCAM data. We do not use these data as they are too sparse (typically $<$5 sources) and provide little constraint on the thermal far-IR/sub-mm, which is our main focus here.\\
The radio fluxes at 1.4\,GHz given in Table~1 are from a VLA survey of the field \cite{ci99}, and errors are 1\,$\sigma$.
\section{Results \& Analysis}
\subsection{Assembling the multiwavelength data}
The 1999 data presented here were previously discussed in Scott et al.~(2000)\footnote[1]{Note that we use the naming scheme of Dole et al. (2001), which differs from the earlier convention used. In particular N1-038, N1-061, and N1-063 from Scott et al.~(2000) correspond to the new N1-040, N1-048, and N1-064.}. However we re-reduced the old data concurrently with the 2001 data in order to ensure uniformity, especially as an upgraded version of {\sc surf} and a new custom-written code were used.  We confirm all previously reported detections (i.e. targets having SNR$\,{>}\,3$). The 2001 data have three unambiguous detections at 450\,$\mu$m (N1-004, N1-024, N1-078). This high detection rate at a difficult band is due to the exceptional atmospheric conditions during our observing run, as well as the far superior performance of the new wide-band filter. In addition there are three new detections at 850\,$\mu$m (N1-001, N1-059, N-078). Note that the few arcsecond pointing uncertainty (see Section~2.2) has only a small effect on the long wavelength data, but may be a significant reason for the apparently missing 450\,$\mu$m flux (where the beam FWHM is only $\sim$\,8\,arcsec) in sources where one would expect to find some (e.g. N1-059). 
\begin{table*}
\centering
\begin{minipage}{110mm}
\caption{Multiwavelength data for our sample. All errors are 1\,$\sigma$ estimates.}
\begin{tabular}{lcrrrrr}
Source & $K$$^a$ & $S_{100}$$^b$ & $S_{170}$$^c$ & $S_{450}$$^d$ & $S_{850}$$^d$ & $S_{1.4\rm{GHz}}$$^e$  \\
 & mag & mJy & mJy & mJy & mJy & mJy \\
 & & & & &  &  \\
N1-001 & 12.4$\pm$0.1$\phantom{^\dagger}$ & 430$\pm$87 & 597$\pm$72 & --3.0$\pm$14.0 & 6.1$\pm$1.6 & 0.74$\pm$0.23  \\ 
N1-002 & 12.7$\pm$0.1$\phantom{^\dagger}$ & 340$\pm$121 & 544$\pm$69 & 14.4$\pm$12.4 & 4.4$\pm$1.1 & 0.64$\pm$0.04  \\ 
N1-004 & 12.4$\pm$0.0$^{\dagger}$ & 300$\pm$73 & 391$\pm$58 & 32.5$\pm$7.2 & 3.6$\pm$1.4 & 0.88$\pm$0.13  \\ 
N1-007 & 13.2$\pm$0.1$^{\dagger}$ & 480$\pm$73 & 338$\pm$54 & 23.4$\pm$8.1 & 4.4$\pm$1.6 & 1.04$\pm$0.12  \\ 
N1-008 & 14.2$\pm$0.1$^{\dagger}$ & 160$\pm$73 & 335$\pm$54 & 25.9$\pm$14.0 & 1.9$\pm$1.1 & 2.98$\pm$0.04  \\ 
N1-009 & 12.0$\pm$0.0$\phantom{^\dagger}$ & 310$\pm$58 & 313$\pm$52 & 10.6$\pm$7.6 & 3.5$\pm$1.5 & 1.15$\pm$0.11  \\ 
N1-010 & 13.0$\pm$0.0$\phantom{^\dagger}$ & 360$\pm$99 & 309$\pm$52 & 15.2$\pm$10.8 & 1.8$\pm$1.4 & 1.05$\pm$0.20  \\ 
N1-012 & 13.9$\pm$0.2$\phantom{^\dagger}$ & 320$\pm$122 & 302$\pm$51 & 9.2$\pm$10.0 & 1.5$\pm$1.6 & 0.31$\pm$0.07  \\ 
N1-013 & 16.8$\pm$0.1$^{\dagger}$ & 350$\pm$75 & 294$\pm$51 & 18.8$\pm$9.9 & 0.0$\pm$1.5 & 0.52$\pm$0.15  \\ 
N1-015 & 14.8$\pm$0.1$\phantom{^\dagger}$ & 230$\pm$41 & 294$\pm$51 & --3.4$\pm$7.7 & 1.4$\pm$1.6 & 0.52$\pm$0.07  \\ 
N1-016 & 13.2$\pm$0.1$\phantom{^\dagger}$ & 360$\pm$92 & 289$\pm$50 & 34.8$\pm$16.7 & 1.5$\pm$1.2 & 1.55$\pm$0.13  \\ 
N1-024 & 14.2$\pm$0.0$^{\dagger}$ & ${<}$\,147 & 266$\pm$49 & 32.3$\pm$7.5 & 2.9$\pm$1.3 & 0.75$\pm$0.02  \\ 
N1-029 & 14.3$\pm$0.1$\phantom{^\dagger}$ & 340$\pm$83 & 229$\pm$46 & 20.0$\pm$14.2 & 0.5$\pm$1.7 & 0.69$\pm$0.05  \\ 
N1-031 & 13.5$\pm$0.1 & 110$\pm$78 & 225$\pm$46 & 9.2$\pm$13.2 & 1.9$\pm$1.1 & 0.43$\pm$0.06  \\ 
N1-032 & 18.5-19.5$^{\dagger}$ & 220$\pm$53 & 224$\pm$46 & 14.9$\pm$7.7 & 1.3$\pm$1.4 & 0.21$\pm$0.05  \\ 
N1-034 & 19.3$\pm$0.8$^{\dagger}$ & 270$\pm$71 & 221$\pm$46 & 95.1$\pm$27.5 & 1.3$\pm$1.3 & 0.33$\pm$0.07  \\ 
N1-039 & 15.8$\pm$0.2$\phantom{^\dagger}$ & 230$\pm$ 0 & 205$\pm$44 & 10.9$\pm$86.8 & --0.1$\pm$2.3 & 0.58$\pm$0.07  \\ 
N1-040 & 19.4$\pm$0.6$^{\dagger}$ &   $<$\,153 & 205$\pm$44 & 29.2$\pm$20.5 & 5.4$\pm$1.1 & 0.33$\pm$0.03  \\ 
N1-041 & 14.7$\pm$0.1$\phantom{^\dagger}$ & 150$\pm$55 & 204$\pm$44 & 20.4$\pm$156.7 & --0.1$\pm$2.5 & 0.76$\pm$0.06  \\ 
N1-045 & 14.4$\pm$0.0$^{\dagger}$ & $<$\,165 & 198$\pm$44 & 15.3$\pm$8.3 & 3.0$\pm$1.4 & 0.43$\pm$0.06  \\ 
N1-048 & 19.3$\pm$0.9$^{\dagger}$ & $<$\,165 & 192$\pm$44 & 15.5$\pm$11.6 & 4.2$\pm$1.1 & 0.37$\pm$0.05  \\ 
N1-056 & 15.9$\pm$0.2$\phantom{^\dagger}$ & 160$\pm$71 & 179$\pm$43 & --8.6$\pm$8.4 & 0.0$\pm$1.6 & 0.24$\pm$0.02  \\ 
N1-059 & 20.4$\pm$0.9$^{\dagger}$ & 230$\pm$66 & 175$\pm$42 & 6.9$\pm$34.5 & 6.4$\pm$1.9 & 0.60$\pm$0.06  \\ 
N1-064 & 18.2$\pm$0.3$^{\dagger}$ & 260$\pm$79 & 166$\pm$42 & 35.2$\pm$13.9 & 5.1$\pm$1.2 & 0.23$\pm$0.04  \\ 
N1-068 & 15.3$\pm$0.1$^{\dagger}$ & 320$\pm$95 & 165$\pm$42 & 15.1$\pm$7.6 & 2.2$\pm$1.4 & 0.44$\pm$0.05  \\ 
N1-077 & 15.5$\pm$0.2$\phantom{^\dagger}$ & 200$\pm$89 & 159$\pm$41 & 5.9$\pm$7.3 & 1.1$\pm$1.3 & 0.40$\pm$0.10  \\ 
N1-078 & 18.0$\pm$0.4$^{\dagger}$ & 240$\pm$63 & 158$\pm$41 & 35.2$\pm$8.7 & 5.7$\pm$1.3 & 0.24$\pm$0.04  \\ 
N1-083 & 15.4$\pm$0.2$\phantom{^\dagger}$ & $<$\,195 & 150$\pm$41 & 16.2$\pm$16.0 & 0.7$\pm$1.2 & 0.55$\pm$0.03  \\ 
N1-101 & 15.2$\pm$0.2$\phantom{^\dagger}$ & 210$\pm$73 & 136$\pm$40 & 19.8$\pm$7.5 & 0.9$\pm$1.5 & 0.39$\pm$0.05  \\ 
N1-153 & 15.5$\pm$0.2$\phantom{^\dagger}$ & 140$\pm$58 & 103$\pm$37 & 9.6$\pm$15.3 & --0.2$\pm$1.0 & 0.24$\pm$0.03  \\ 
N2-013 & --- & 310$\pm$75 & 244$\pm$53 & 23.5$\pm$15.9 & 3.5$\pm$1.4 & 0.30$\pm$0.07  \\ 
 & & & &  &  & \\
\end{tabular}

\medskip

$^a$ These were obtained from our UKIRT sample (marked by $^\dagger$, see also Fig.~2) when available and from the 2MASS catalogue otherwise. Those fainter than $K\sim14$ were estimated directly from the 2MASS images. \\
$^b$ {\sl IRAS }100\,$\mu$m fluxes obtained using {\sc xscanpi} (see text). \\
$^c$ ISOPHOT 170\,$\mu$m data from Dole et al. (2001). \\
$^d$ SCUBA 450\,$\mu$m and 850\,$\mu$m fluxes from this work. \\
$^e$ VLA 21\,cm fluxes from Ciliegi et al. (1999). \\
\end{minipage}
\end{table*}
Our results for the entire sample are presented in Table~1, which, in addition to the sub-mm data, includes near-IR, far-IR, and radio data. \\
The sample as a whole is strongly detected, with average fluxes of $\left\langle S_{850}\right\rangle$=2.6$\pm$0.2\,mJy and $\left\langle S_{450}\right\rangle$=16.7$\pm$1.9\,mJy. This is indicative of the fact that although most sources do not clear the $3\sigma$ detection threshold, there is useful statistical information in the flux data (far in excess of what would be expected for mere sky fluctuations). Thus we treat these fluxes directly, along with their uncertainties, rather than just using the upper limits. Here we focus on the general properties of the sample which are revealed by the addition of the sub-mm and near-IR data. Physical properties for individual sources will be explored elsewhere \cite{lg03}.\\  
Two possible sources of uncertainty in comparing the far-IR and SCUBA fluxes are: 1) the different sizes of the {\sl ISO}(about 90\,arcsec) and SCUBA (about 15\,arcsec) beams, and thus the possibility of multiple sources lying within the beam -- this is also complicated by the issue of clustering, and possibly lensing; and 2) the relative proximity of the bulk of our sources and thus the possibility of at least a few being extended beyond the SCUBA beam (see for example N1-004 and N1-007 in Fig.\,2). This is especially severe for the $\sim$\,8\,arcsec 450\,$\mu$m beam. In general, however, images at other wavelengths, are insufficient to determine precisely how much flux is being lost in this way, as it is not clear how concentrated the sub-mm emission would be, or how it is distributed in comparison with other wavebands. In any case, the sub-mm flux estimates are likely to be less accurate in photometry mode if the source size is on the order the beam size. These effects, of course, become more important with increasing $S_{\rm{170}}$. \\
Although the low sub-mm flux of most of our sources is consistent with local star-forming galaxies, there is the additional concern about the possibility of cirrus contamination of the FIRBACK sample. Previous studies have discussed this issue extensively \cite{lp00,p99}, concluding that the FIRBACK sources are unlikely to suffer from significant cirrus contamination. The N1 field was chosen for this survey due to its low cirrus emission. We can estimate, using the standard formula \cite{hb90}, that the cirrus variance is $\sim$10 times less than the detection cutoff of the FIRBACK survey. Due to the non-Gaussianity of the cirrus fluctuations \cite{g92} some outliers are still possible, but these would have to reside in very cold cores which have not been revealed in deep searches for CO emission. Moreover, the far-IR slope of cirrus knots is known to be much steeper than that of sources. A study, whose source selection is similar to FIRBACK's \cite{j00}, makes use of this in concluding that cirrus contamination is highly unlikely. In agreement with that study, our sources show a fairly flat $S_{\rm{100}}/S_{\rm{170}}$ ratio. Finally, all our FIRBACK sources have a radio counterpart, further reducing the chance that cirrus contamination is relevant. 
\subsection{Linear Correlations}
Fig.~3 shows three illustrative projections for our data, using the 170\,$\mu$m, 450\,$\mu$m, 850\,$\mu$m and radio fluxes. For each plot a fit is made to all points using a straight line passing through the origin, i.e. $y=mx$. Then the $>2\sigma$ outliers are excluded, a new fit is made and is plotted along with the rms scatter. Table~2 shows the results. Since in most cases, the errors in the two directions are of comparable size, rather than use the conventional 1-D $\chi^2$, we minimize a statistic which combines errors in both directions: 
\begin{equation}
\chi^{2}_{\rm{2D}} = \sum_{i=1}^N\frac{(y_i-mx_i)^2}{(\sigma_{yi}^2+m^2\sigma_{xi}^2)},
\end{equation}
where $\sigma_x$ and $\sigma_y$ are the two errors. 
\begin{figure}
\centering
\vspace*{22cm}
\leavevmode
\includegraphics{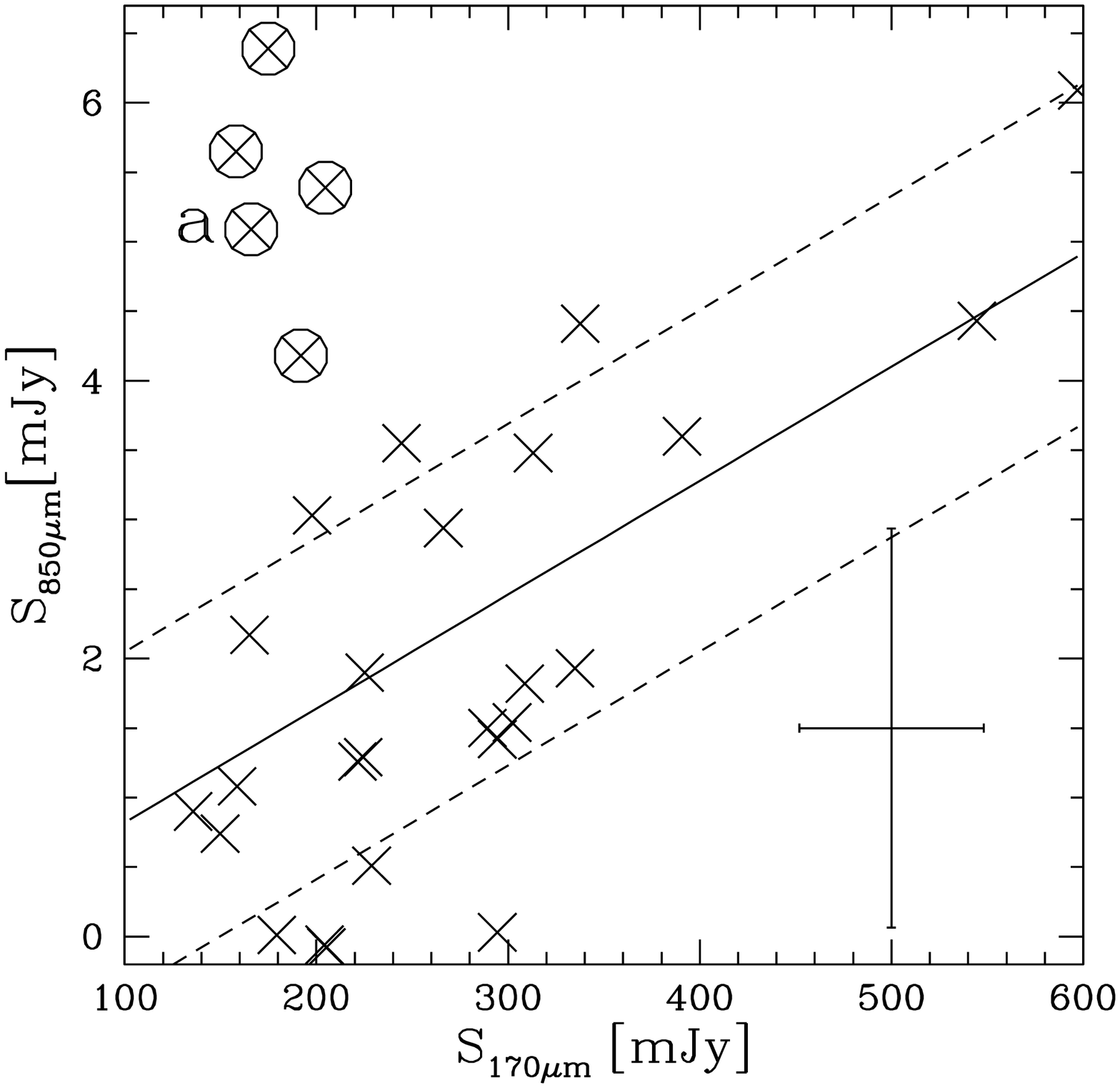}
\includegraphics{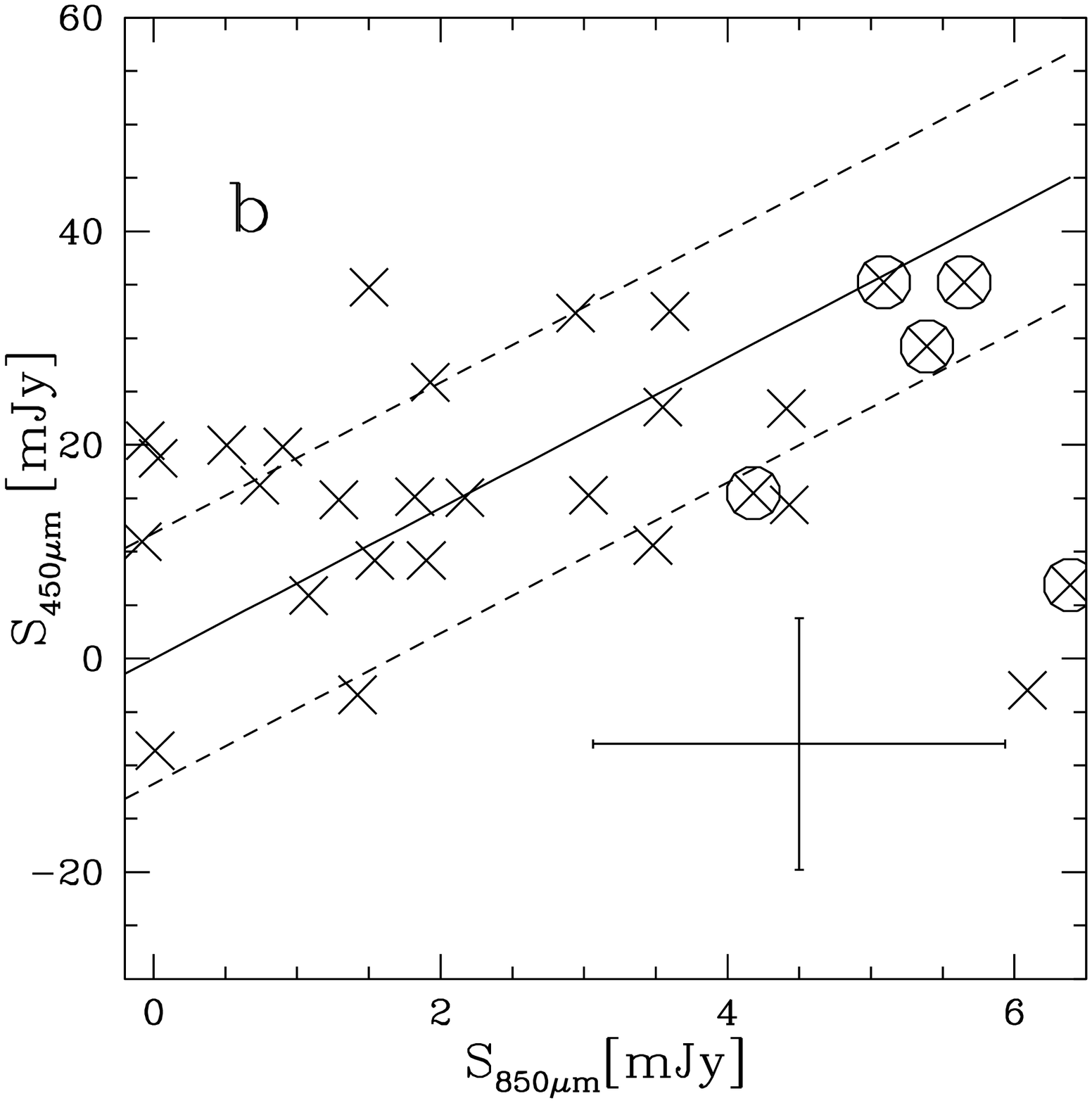}
\includegraphics{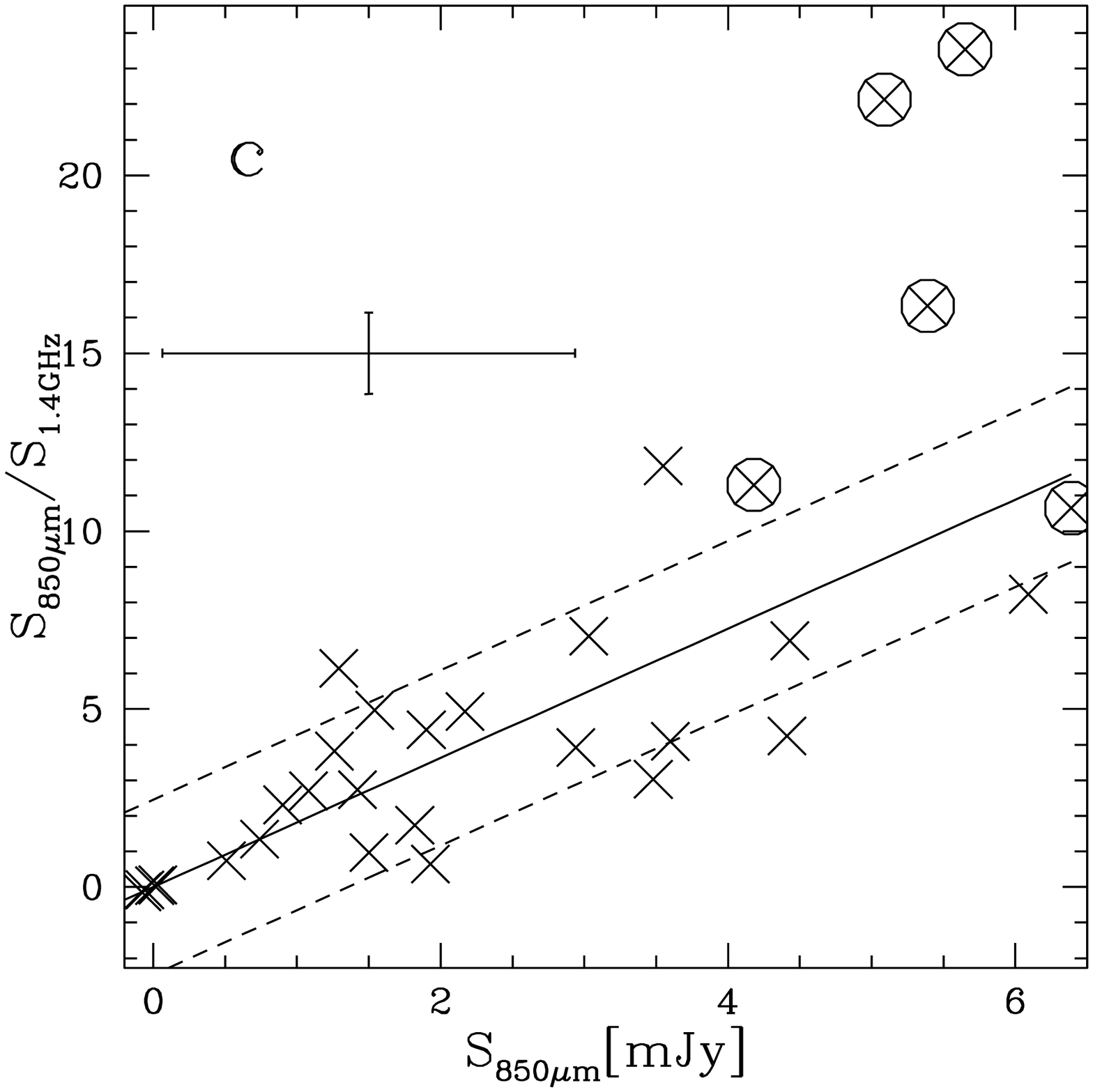}
\caption{Multiwavelength projections of our data.The solid lines are the $y=mx$ fits obtained by minimizing $\chi_{\rm{2D}}^{2}$ equation~(1). The dashed lines are the $\pm 1\sigma$ scatter of the points around the lines, with the fit parameters given in Table~2. The error bars shown are representative for our data. The 5 outliers in panel (a) are indicated with the same symbols in the other two panels.}
\end{figure}
In general, it appears that the typical uncertainties of our sub-mm data points are larger than the 1\,$\sigma$ scatter of the fits. We test this explicitly for the $S_{\rm{450}}$ vs. $S_{\rm{850}}$ plot via Monte Carlo simulations of the data, thereby estimating the probability of the $\chi^{2}_{\rm{2D}}$ obtained to be $\sim2\%$. This implies that our errors are overestimated by a factor of $\sim\sqrt{2}$ or else there is some as yet unidentified source of correlation in the errors. We will, however, be conservative and leave the errors as they are, since we cannot properly account for the source of this discrepancy. \\ 
After this cautionary aside, we return to Fig.~3. It is organized such that the order roughly follows the main features of the long wavelength SED: the top plot largely tracks the location of the thermal peak; the middle corresponds to the sub-mm slope; and the bottom plot traces the trough between the thermal and non-thermal emission. \\
First, we will concentrate on the $S_{850}$ vs.~$S_{170}$ plot (Fig.~3a). We immediately notice that about 5 sources (the circled crosses) occupy a locus $>2\sigma$ away from the best-fit line. Assuming a grey-body model, these sources are either at higher redshift or lower temperature than the rest of the FIRBACK 170\,$\mu$m sources (or a combination of both).\\
To understand which, we turn to the $S_{450}$ vs. $S_{850}$ plot (Fig.~3b). Here we are mainly exploring the slope of the spectrum in the sub-mm. 
What we notice in this plot is that there is not a population of outliers as there was in the $S_{850}$ vs. $S_{170}$ plot (the only outliers are N1-001, N1-034, and N1-059 which all have sub-mm slopes which alone cannot be fit by any sensible dust/redshift combination and thus we assume suffer from some systematic effect such as discussed in section~3.1). This common distribution implies that a single [$\beta,T/(1+z)$] combination describes the sample reasonably well. The lack of the outliers, from the $S_{850}$ vs. $S_{170}$ plot (Fig.~3a), means that the 5 outliers' (N1-040, N1-048, N1-064, N1-059, and N1-078) location in Fig.~3a is most likely due to having somewhat higher $z$ rather than a significantly different SED shape. In addition, their redshifts can only be as high as $z\sim$1, corresponding roughly to the shift between 170\,$\mu$m and 450\,$\mu$m (a higher redshift, much warmer population would have to be very finely tuned to still fall on the same distribution, which seems improbable). This in turn roughly constrains the temperature of the sources (by Wien's law) to less than about 40\,K. \\
We now turn to Fig.~3c, the 850\,$\mu$m/1.4\,GHz vs. 850\,$\mu$m comparison, where the previous outliers return again (except for N1-059 which has unusually high radio emission, discussed later), confirming that indeed redshift, rather than dust properties is the main difference between them and the rest of the sample.
\subsection{Sub-mm/Radio redshifts}    
As a tracer of the trough between the thermal and non-thermal emission of galaxies, $S_{850}/S_{1.4\rm{GHz}}$ is commonly used as a redshift indicator \cite{cy00}. It is, however, degenerate in dust properties for galaxies cooler than about 60\,K \cite{bl02}. Since we are clearly in that regime (see previous section), we have investigated using this relation calibrated on samples with different selections. 
The first parameterization we used was the Carilli \& Yun relation (Carilli \& Yun~2000, CY hereafter) which is based on {\sl IRAS}-selected and somewhat more radio-loud galaxies, with likely a higher fraction of AGN. The second parameterization we used was that from  Dunne, Clements \& Eales (Dunne, Clements \& Eales~2000, DCE hereafter) which is based on the SLUGS (SCUBA Local Universe Galaxy Survey) sample \cite{de01} which is essentially an {\sl IRAS}-bright local selection (unlike ours).\\
In general, such relations provide too weak a constraint on individual redshifts locally, due to the large scatter in intrinsic galaxy properties. However, they still confirm our prior selection of N1-040, N1-048, N1-064, N1-059, and N1-078 as likely being at somewhat higher redshifts than the rest of the galaxies in our sample (based on both relations giving a redshift $>0.4$ for all of these sources). The only other source which satisfies this is N2-013. We return to it in section~3.6, when we also discuss different redshift estimators.
\subsection{Sub-mm vs. near-IR}
Here we examine the correlation between the $K$ magnitudes and 850\,$\mu$m fluxes of our sources. The $S_{\rm{850}}$ flux density by itself is not a good redshift indicator due to its $k$-correction behaviour.  It is however a good luminosity tracer.  The $K$ magnitude (in the rest frame) is also a luminosity indicator, as it is 10 times less obscured than the optical. The $S_{\rm{850}}$-to-$K$ magnitude relation can be used as a redshift indicator \cite{b99,d02} because the rest-frame shorter wavelengths (that are much more dust obscured) move with increasing redshift into the observer-frame near-IR. Thus the $K$ magnitude for a given sub-mm flux is dependent on both redshift and dust obscuration. Fig.~4 plots the flux at 850\,$\mu$m against $K$ magnitude, where the higher-$z$ sources clearly populate a different locus from the nearby galaxies. They are roughly 3\,$\sigma$ removed from the best-fit relation for the other sources, with a distinct gap between the two groups (apart from a couple of sources which we discuss later). The gap is more pronounced than in Fig.~3, as here the $(1+z)/T_{\rm{d}}$ degeneracy is somewhat broken. \\ 
\begin{table}
\centering
\caption{Results for the linear fits to the data}
\begin{tabular}{@{}cccc@{}}
Relation & m & rms & $\chi^2$  \\
850\,$\mu$m vs. 170\,$\mu$m & 0.01 & 1.23 & 18.07 \\ 
450\,$\mu$m vs. 850\,$\mu$m & 7.05 & 11.74 & 13.95 \\ 
850\,$\mu$m/1.4\,GHz vs. 850\,$\mu$m & 1.82 & 2.46 & 12.19 \\ 
\end{tabular}
\end{table}
A redshift relation based solely on the $K$ magnitude is bound to be degenerate in some other galaxy properties (amount of dust, luminosity etc.), and can only work for a very homogeneous sample of galaxies (see e.g. Willot et al.~2002). The addition of sub-mm flux improves this relation, since the $K$-band is dust absorption attenuated (specifically for distant, luminous galaxies), whereas the sub-mm flux arises from dust emission, and thus the combination of the two will somewhat break the dust degeneracy. This diagnostic of both the absorption and emission spectrum allows for a more robust redshift indicator over different galaxy types. What is robustly clear is that, in general, objects which are detected at 850\,$\mu$m and are faint at $K$-band are at higher redshift. In Fig.~4 we find the same handful of outliers as in Sections 3.2 and 3.3.
\subsection{SED fits}
We now fit single grey-body SEDs to the 170\,$\mu$m, 450\,$\mu$m, and 850\,$\mu$m fluxes of each source (see Fig.~5). We assume optically-thin sources; the effect of including a non-negligible $\tau$ in the fits is to suppress the peak with respect to the Rayleigh-Jeans tail, such that the best-fit dust temperature inferred will be 10--20$\%$ higher than otherwise \cite{bl02}. This is therefore not an important effect for sub-mm spectra dominated by single temperature dust emission (to the precision of our fits of three points).  In order to avoid stretching these assumptions too much, we only use the 170\,$\mu$m, 450\,$\mu$m, and 850\,$\mu$m fluxes in the fits, where the SED is dominated by the coldest significant dust component. \\
We fit for two parameters -- the overall normalization, which gives the luminosity (if the redshift is known), and the wavelength shift, which is proportional to $(1+z)/T_{\rm{d}}$. The best fit $\chi^2$ and fit parameters were obtained with the {\sc pikaia} genetic algorithm \cite{c95}.
With only 3 points for each galaxy, we could not also fit for the emissivity index $\beta$, and therefore it was held constant. However, we investigated how different values of $\beta$ (1.3, 1.5, 1.7, and 1.9) would affect the quality of fits for the sample as a whole. Values of $\beta=1.5-1.7$ are far better than either smaller or larger values, for which a third of the sample has a $\chi^2>2$. The sources, with a poor fit in all cases are N1-001, N1-002, N1-015, N1-034, and N1-056. For the lowest $\beta$ tested, 5 additional sources show a poor fit: N1-008, N1-009, N1-012, N1-013, N1-016 (note that these are all among the brighter {\sl ISO} sources in our sample). On the other hand, the higher, $\beta=1.9$, value, provides a poor fit for all the sources we singled out in earlier sub-sections as potentially being at higher redshifts (see Fig.~5). For local ULIGs, previous studies \cite{de01,kl01} already showed this trend, with lower $\beta$ providing a better fit when single temperature grey-bodies are used, whereas generally a $\beta\sim2$ is inferred for multi-temperature fits. \\
Notice that in Fig.~5 we also show the 100\,$\mu$m point, although it is not included in the fits. Fitting to all four points results in poor fits either on the shorter or longer wavelength ends, depending on the choice of $\beta$. To fit this wider range presumably requires a multi-component model, but the data do not warrant a detailed investigation of more complicated SEDs. \\
\begin{figure}
\centering
\vspace*{8.0cm}
\leavevmode
\includegraphics{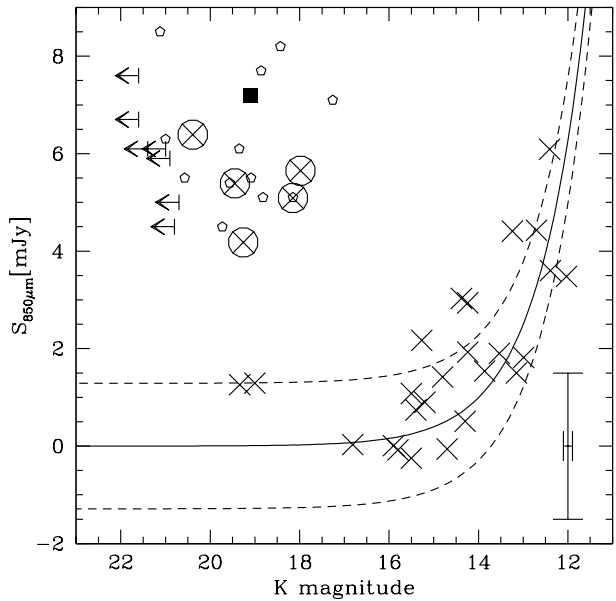}
\caption{
$S_{850}$ vs $K$ magnitude for our sample (crosses, with the 5 outliers also encircled). The filled square and $K$-band upper limits are from the SCUBA lens survey (Smail et al.~2001), while the pentagons are from the UK 8\,mJy survey (Ivison et al.~2002). Our high-$z$ candidates populate a similar region to these other SCUBA survey sources. The solid line is a $\log$--$\log$ fit to the crosses only, with the dashed lines being the $\pm 1\sigma$ scatter. The error bar in the lower right-hand corner is a representative one for our measurements.
} 
\end{figure}
\begin{figure}
\centering
\vspace*{9.5cm}
\leavevmode
\includegraphics{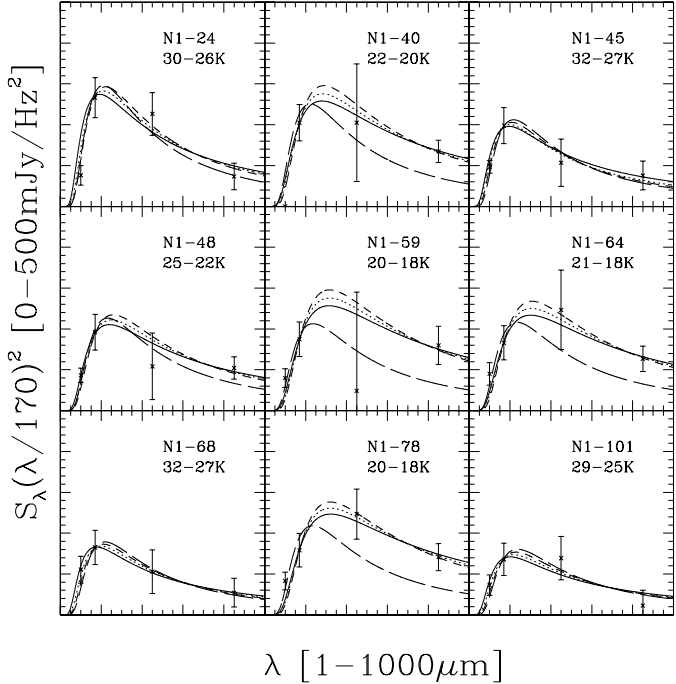}
 \caption{The SED fits for a fraction of our sample where the fits suggest a non-zero redshift. The axes scales are linear with the $x$,$y$-ranges in all panels being the same (shown in the labels), and the sources are arranged by decreasing 170\,$\mu$m flux. The {\sl IRAS} 100\,$\mu$m points included in the plots are not used in the fit. For the sake of clarity, we rescale the flux via $(\lambda$/170\,$\mu$m)$^2$. This was necessary as there is typically an order of magnitude difference between the the far-IR and sub-mm fluxes, and in this way the sub-mm, which is the focus of this work, is emphasized. The solid line corresponds to $\beta$=1.3, the dotted line to $\beta$=1.5, the short-dashed line to $\beta$=1.7, and the long-dashed line to $\beta$=1.9. The truncated temperatures shown are in units of $\frac{T}{(1+z)}$ and correspond to the $\beta$=1.5 and 1.7 cases respectively. See text for an explanation of this choice as well as further discussion on the fits.}
\end{figure}
It thus appears that it is reasonable to model our data as a single temperature grey body in the limited spectral range considered (i.e. 170\,$\mu$m$\rightarrow$850\,$\mu$m), provided we explore the effects of varying the parameters within the accepted range we found here. Of course this model will be incorrect in detail for the entire thermal spectrum, as has been shown previously by several authors \cite{de01,bl02,st00,kl01}. In general, any grey-body model, even with multiple components, remains only an approximation to the underlying far more complex dust properties. We also caution that the dust properties are merely phenomenological, corresponding to emission-weighted averages, and depending on the wavelength range chosen.  Care should therefore be taken in interpretting any dust temperatures physically.
\subsection{Redshifts, and Luminosities}
Now we wish to examine some of the physical characteristics of the non-zero redshift sources in our sample (because they are the best constrained through their sub-mm fluxes). Since, fundamentally, the dust spectrum/distance degeneracy is still present, derivations of properties such as luminosity have an intrinsic uncertainty. However, we can provide physical parameter estimates for what we consider to be reasonable ranges of redshifts (for the likely higher redshift sources only) in order to obtain a handle on the physical nature of our sources for comparison purposes.\\
A few of our optically-fainter sources have available spectroscopic redshifts obtained from Keck and Palomar spectroscopy. These are N1-045($z$=0.25); N1-068($z$=0.22); N1-039($z$=0.27); N1-008($z$=0.27); N1-040($z$=0.45); and N1-064($z$=0.91). The details of the spectroscopic observations, and the obtained redshifts are discussed elsewhere \cite{c03,c02}. These provide valuable checks of our redshift estimates. However, the spectroscopic redshifts only exist for a fraction of our sample, having strong selection effects.  Therefore, we do not make more direct use of them as we wish to treat our sample in a uniform way. The details of the spectroscopic observations, and the obtained redshifts are discussed elsewhere \cite{c03}.    \\  
For definitiveness we focus on the ($\beta$, $T_{\rm{d}}$) combinations which agree with the SED fit results (see section~3.5), with the redshifts estimated from the sub-mm/radio relations (section~3.3), and with expectations from other studies (see section~4). These are (1.7, 30\,K), and (1.5, 40\,K). Table~3 shows the redshift ranges, along with luminosities\footnote{We solve for the total far-IR luminosity simply by integrating our fitted grey-body SEDs. The key mathematical step is the integral $\int^{\infty}_{0}x^{s-1}(e^x-1)^{-1}dx=\Gamma(s)\zeta(s)$.}, and correspoinding distances for the higher redshift candidates in our sample. The ranges illustrate how the values vary with input parameters, and also give an indication of the spread of reasonable values. The sources in the upper half of the table are our $S_{\rm{850}}>3\sigma$ detections (excluding N1-001, and N1-002 which are consistent with being at zero redshift). They are consistent with $z\sim0.4-0.9$ ULIGs, based on the average luminosity of this sub-sample being above $10^{12}\rm{L}_{\odot}$. The likely lowest luminosity (and redshift) source of this set is N1-048, which may be only a Luminous Infrared Galaxy (LIG), usually defined as $<10^{12}\rm{L}_{\odot}$, depending on the exact dust parameters. The bottom half of Table~3 contains five LIG candidates (N1-101, N2-013, N1-024, N1-068, and N1-045) based on the range of ($\beta$,\,$T_{\rm{d}}$) we considered. 
\begin{table}
\begin{center}
\caption{Fit results for all the higher-$z$ sources$^a$}
\begin{tabular}{@{}rccc}
Source & $z$ & $\log(L)$ [L$_{\odot}$] & $D_{\rm{L}}$ [Mpc]\\
          N1-078   &  0.66--1.00  & 12.1--12.6  &  3700--6100  \\
          N1-059   &  0.64--0.96  & 12.2--12.6  & 3500--5900   \\
          N1-064  & 0.58--0.89(0.91)  & 12.1--12.6  & 3200--5300   \\
          N1-040  & 0.49--0.76(0.45)  & 12.0--12.5 & 2600--4400 \\
          N1-048  & 0.33--0.56  & 11.7--12.3  & 1600--3000  \\
\hline
	  N1-101  & 0.18--0.35 & 11.0--11.8 & 810--1700 \\
	  N2-013 &  0.15--0.30 & 11.1--11.9 &  684--1500 \\
 	  N1-024 & 0.14--0.30 & 11.1--12.0 & 621--1400 \\
          N1-068  & 0.08--0.22(0.22) & 10.4--11.5  & 330--1000 \\
          N1-045  & 0.08--0.22(0.25) & 10.5--11.6  & 320--1000 \\
\end{tabular}

\smallskip
\end{center}

$^a$ The ranges correspond to different ($\beta,T_{\rm{d}}$) combinations, the first being (1.7,\,30\,K) the second (1.5,\,40\,K).  The redshifts in brackets are measured spectroscopically. The top half of the table has the $>3\sigma$ detections, while the bottom half of the table has the other sources from the sample with redshifts from the fits having lower limits greater than zero. The sources are arranged by decreasing redshift. 
\end{table}
When comparing the redshift ranges obtained using the above rather approximate method against the known spectroscopic redshifts, we find fairly good agreement between them. This means that the assumptions made were probably reasonable. However, the two ULIGs with spectroscopic redshifts have measured values at the upper and lower end of the estimated ranges, perhaps suggesting a slightly different temperature, at least in some cases. \\
The only other exceptions here were N1-008 and N1-039, since, with their $z_{\rm{spec}}=0.27$, they should have been grouped with the others above. 
There is no reason to suspect intrinsically different dust properties in either case. It is worth noting that there are two radio sources within the N1-008 {\sl ISO} beam. The missing sub-mm flux of N1-039 may be related to the fact that it is one of the sources which were solely observed during the poorer weather conditions of the May, 2001 observing run -- rms = 2.3\,mJy (see Table~1), compared to a mean rms of 1.5\,mJy for the sample.\\
Comparing with the redshift estimates derived from the far-IR/radio correlations (Section~3.3), we find broad agrement. The DCE relation generallly gives good overlap with the other methods of estimating redshift, while the CY relation gives redshifts which are somewhat high.  This is because the effective temperature in the DCE relation is more appropriate for our sample.  
\subsection{Star Formation Rates}
The first star formation rate (SFR) estimator we consider is based on the FIR luminosities. This uses the fact that the thermal dust emission is reprocessed stellar light which was absorbed primarily in the UV (i.e. where young stars are the main contributors). For a discussion of the uncertainties associated with this and other SFR estimators see e.g. Schaerer et al.~(1999). Essentially, the problem is to find a universal calibration given the vastly varying conditions (e.g. different dust properties, and contribution of cirrus, stars,  or AGN to the thermal spectrum) from galaxy to galaxy. Here we adopt a form, which is based on stellar evolution models using a Salpeter IMF \cite{char02}: 
\begin{equation}
\frac{\rm{SFR_{\rm{FIR}}}}{[\rm{M}_{\odot}\rm{yr}^{-1}]}=1.7\times10^{-10}\frac{L_{\rm{FIR}}}{\rm{L}_{\odot}}.
\end{equation}
 The second method is based on the radio continuum, where the idea is essentially the same as above, but use is also made of the well known far-IR/radio correlation (see Section~3.3). Here we use an empirical relations based on {\sl IRAS} galaxies with radio observations \cite{y01}:
\begin{equation}
\frac{\rm{SFR_{\rm{rad}}}}{[\rm{M}_{\odot}\rm{yr}^{-1}]}=(5.9\pm1.8)\times10^{-22}\frac{L_{\rm{1.4GHz}}}{[\rm{WHz}^{-1}]}
\end{equation}
The radio luminosity is then obtained via
\begin{equation}
\log L_{\rm{1.4GHz}}=20.08+2\log D_{\rm{L}}+\log S_{\rm{1.4GHz}},
\end{equation}
where $L$ is in $\rm{W Hz}^{-1}$, $D_{\rm{L}}$ is in Mpc and $S_{\rm{1.4GHz}}$ is in Jy.\\
\begin{table}
\centering
\caption{Estimating the Star Formation Rates$^a$.}
\begin{tabular}{@{}rccc}
Source & $z$ & SFR$_{\rm{rad}}$[M$_{\odot}\rm{yr}^{-1}$]  & SFR$_{\rm{FIR}}$[M$_{\odot}\rm{yr}^{-1}$] \\
         N1-078   &  0.66--1.00  & 230--640 & 240--730 \\
          N1-059   &  0.64--0.96  & 530--1460  & 250--740  \\
          N1-064  & 0.58--0.89  & 170--460 & 200--620  \\
          N1-040  & 0.49--0.76  & 160--450 & 170--580 \\
          N1-048  & 0.33--0.56  & 70--240 & 80--310 \\
\hline
	  N1-101  & 0.18--0.35 & 20--80 & 20--100 \\
	  N2-013 &  0.15--0.30 & 10--50 & 20--150  \\
	  N1-024 & 0.14--0.30 & 20-110 & 20--160 \\
          N1-068  & 0.08--0.22 & 3--30  & 4--60 \\
          N1-045  & 0.08--0.22 & 3--30  & 5--70 \\

\end{tabular}

\smallskip

$^a$ This table is arranged in an same way as Table~4.
SFR$_{\rm{rad}}$ uses the relation of equation~(3), while SFR$_{\rm{FIR}}$ uses equation~(2).
\end{table}
We present the results in Table~4.
The two estimators agree with each other reasonably well (as expected for star-forming galaxies). The strongest deviation is observed for N1-059. This source has a greater radio flux by a factor of roughly $\sim2$ from its peers, which may be an indication of AGN contribution. In general, estimates for lower luminosity sources are more affected by the various systematic uncertainties discussed above. The conclusions are: that the higher-$z$ candidates have $L\sim10^{12}$\,L$_{\odot}$ and SFRs of typically a few hundred M$_{\odot}$yr$^{-1}$; and that the other galaxies with estimated redshifts above 0 have $L\sim10^{11}$\,L$_{\odot}$ and SFRs of typically a few tens of M$_{\odot}$yr$^{-1}$. The rest of our sample, which are more nearby, have lower $L_{\rm{FIR}}$ and SFRs, consistent with being more normal star-forming galaxies.
\section{Discussion}
\subsection{Summary of spectral properties}
The multiwavelength photometric analysis of the sample of galaxies presented in the previous sections provides us with an insight into the brightest contributors to the CIB. Our sample also holds information on galaxy evolution roughly in the range $z\sim0-1$. \\
The series of scatter plots in Fig.~3 showed that a group of 5 sources (N1-040, N1-048, N1-059, N1-064, and N1-078) stand out from the rest in the far-IR/sub-mm, and sub-mm/radio projections. Their position in the far-IR/sub-mm plot can be explained either by their being colder, or at somewhat higher redshift than the rest of the sample (the $T_{\rm{d}}/(1+z)$ degeneracy). However, when the sub-mm slope ($S_{450}/S_{850}$) alone is examined, these sources do not stand out as one might have expected if their intrinsic SED shapes were substantially different from the rest of the sample. Thus, we assume an approximately constant SED shape across the sample, and arrive at a combined best-fit single grey-body (from the $S_{170}/S_{450}$ and $S_{450}/S_{850}$ slopes) with $\beta\simeq1.5$ and $T_{\rm{d}}/(1+z)\simeq$\,30\,K. \\
When the sub-mm/radio relation is examined, the same group of 5 stands out again, with two redshift estimators agreeing with their being at higher ($\sim$0.4--1.0) redshifts. Due to uncertainties in the spectral indices, and more importantly scatter in intrinsic galaxy properties these indicators have $\Delta z\sim 0.5$. The DCE \cite{dce00} relation seems to provide the best-fit to the available spectroscopic redshifts, and may be more reliable. The above five sources are the only ones for which this relation gives a redshift $>0.4$\footnote{An additional candidate is N2-013, but on the basis of the SED fits it appears to be a luminous infrared galaxy at perhaps up to $z$\,$\sim$\,0.3, which is consistent with the redshifts provided by the sub-mm/radio relations.}. \\
A different way to look at the data is the sub-mm/near-IR relation (Fig.~4). Here, the segregation of the higher redshift candidates is most pronounced -- more than $2\sigma$ separate each of the above 5 sources from any of the rest of the sample (although the location of N1-032 and N1-034, being faint at $K$ and faint with SCUBA, is poorly understood at this point). This projection has the advantage of sampling two spectral regions with completely different emission mechanisms: thermal dust emission vs. stellar light plus dust attenuation. This means that the $T_{\rm{d}}/(1+z)$ degeneracy is partially broken. This relation would suffer from a completely different set of systematic uncertainties than the radio/sub-mm photometric redshifts. Such an approach was already discussed by Dannerbauer et al. (2002) in the context of their mm-selected galaxies compared with the {\sl IRAS}-selected galaxies of the SLUGS sample \cite{de01}.
However, since no robust $K$--$S_{850}$ relation is known at present, we can do little beyond obtaining a qualitative confirmation of the relative redshifts of our sources.\\  
Finally, we used the knowledge of the general trends in our sample, inferred from the above steps, to attempt to constrain some of the properties of the individual galaxies.
We fitted single, optically-thin, grey-bodies to the 170\,$\mu$m, 450\,$\mu$m, and 850\,$\mu$m points. For the sample as a whole only $\beta\simeq$1.5--1.7 provides a good fit, while 1.3 is a poor fit for many of the probably local sources, and 1.9 is a poor fit for our higher-$z$ candidates. Since the fits only provide $T_{\rm{d}}/(1+z)$, a dust temperature needs to be assumed in order to obtain a redshift. We esimate that an acceptable range is the ($\beta$, $T$) combinations (1.7,\,30\,K)$\rightarrow$(1.5,\,40\,K). These give redshifts which are in reasonable agreement with all relations examined so far (including the DCE, and CY redshift indicators). We used this range to estimate the luminosities and SFRs of the 5 high-$z$ candidates, along with some intermediate sources which are possible LIGs up to $z\sim0.3$, representing the tail of the more local bulk of the sample. \\
Our results, from near-IR to radio, are consistent with having a sample of mostly local galaxies, some slightly higher redshift LIGs and a handful of probable ULIGs at redshifts in the range $0.4<z<1.0$. 
\subsection{Bimodality}
Some of the scatter plots discussed above suggest a bimodality in our sample. Whether our particular observational selection effects result in bimodality in redshift, or just a higher-$z$ tail may be an important point for distinguishing various galaxy evolution models (discussed in the next sub-section). Due to the small size of our high-$z$ candidate sample, it is difficult to test their distribution properties in detail. However, a simple test which we can perform involves comparing the $\chi^2$ resulting from fitting a single line ($y=mx$), against that for fitting two lines, both with zero $y$-intercept. We choose to focus on the $S_{170}$ vs. $S_{850}$ projection as here the bimodality is implied, but is not as clean as in the sub-mm/near-IR relation.  We show the result in Fig.~6. 
\begin{figure}
\centering
\vspace*{8cm}
\leavevmode
\includegraphics{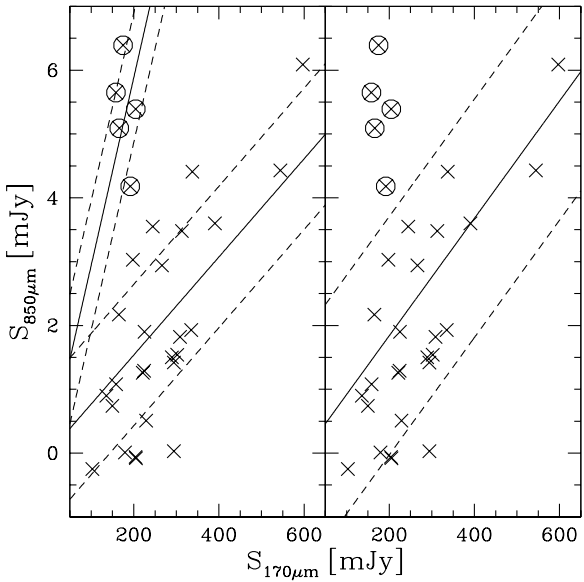}
\caption{Here we test the hypothesis of our sample being bimodal by comparing the $\chi^2$ of a single-line fit for the entire sample (right panel) to a two-line fit for each sub-sample (left panel). The dashed lines are $\pm1\sigma$ where $\sigma$ is the rms scatter in the $y$-direction. Notice that, apart from N1-048, even with the single-line fit to the entire sample, our high-$z$ candidates are $>2\sigma$ away from the best-fit line. See Fig.~3 for representative errorbars.}
\end{figure}
We performed this test simply with the 1D $\chi^2$ and assuming constant error for each source (since they are fairly uniformly distributed in any case). The single line fit results in $\chi^2$ of 108, while the two-line fit results in $\chi^2$ of 37, with $N$=31 (minus either one or two constraints), and for the two-line fit we use the nearest line for each point. The two-line model thus provides an adequate fit to the data, suggesting that each source is drawn from either one population or the other.
This supports the idea that a handful of our sources are a distinct population and lie at $z\sim0.4-0.9$, while most of the sample are at $z\sim0$.   
\subsection{Comparison with evolutionary models}
The sources studied here are a represenetative sample of the brightest $\sim$10$\%$ of sources contributing to the CIB. They thus provide a test of the various evolutionary models abounding in the literature. Models which are consistent with both the observed CIB intensities, and the number counts obtained by various surveys, imply that the majority ($\sim$80$\%$) of the CIB near its peak ($\sim$200\,$\mu$m) will be resolved by sources in the range $0<z<1.5$. The same redshift range sources contribute only $\sim30\%$ of the 850\,$\mu$m background \cite{e02,ce01}. Such models result in a peak of the SFR density at $z\sim1.0$, and then have SFR essentially flat until $z\sim4$. In general $>70\%$ of the star formation takes place in galaxies with $L_{\rm{FIR}}>10^{11}\rm{L}_{\odot}$ \cite{ce01}. \\
From the redshifts we infer for our sample, it seems to span the crucial epoch over which the strongest evolution of the SFR density takes place. In general there is no way to fit the FIRBACK counts without strong far-IR evolution over at least this redshift range. Adopting the starburst template from Lagache et al.~(2002), we see that sources less luminous than about $10^{12}\rm{L}_{\odot}$ fall below the FIRBACK detectability beyond redshift $\sim$0.4. Since the FIRBACK selection excludes normal galaxies beyond $z\sim0.1$ and LIGs beyond $z\sim0.3$, but allows for higher-luminosity sources up to $z\sim1.0$, our mix of normal, starforming galaxies, including a few possible LIGs, and a handful of most likely higher-$z$ ULIGs is in good qualitative agreement with this model.   
The bimodality which this hints at for our selection (and which we appear to observe), is more directly shown by a number of specific models \cite{lag02,c02,ce01,d99,wb00,f01}. An easy way to achieve such a bi-modal distribution is to phenomenologically decompose the luminosity function into a component of normal, quiescent galaxies, and a much more luminous component of ULIGs, and then have the luminous component evolve more strongly than the quiescent one so that it dominates the luminosity funciton by $z\sim1$ \cite{d99,wb00,f01}. This approach was exploited in the context of the FIRBACK sources \cite{lag02}, showing that a double-peaked redshift distribution is predicted, as appears to be confirmed by our data. \\
Another approach to modelling the $N(z)$ distribution is that of Chapman et al.~(2002) where the entire infrared luminosity function is evolved. This combines the colour (i.e.~temperature) distribution of local galaxies with a strong luminosity evolution, such as in Xu (2000), to produce the $N(z)$ distribution for the FIRBACK population. A bimodallity can only be produced here if a bivariate $T_{\rm{d}}$ distribution is used (i.e. by including cold luminous sources), but this is strongly evolution dependent. Discriminating in detail between such models, including issues such as separating density evolution from luminosity evolution, is not currently possible. However, once the full redshift distribution of the FIRBACK sample is obtained, such discrimination may be feasible.\\
The FIRBACK selection allows us to investigate the range 0\,$\stackrel{<}{_{\sim}}$\,$z$\,$\stackrel{<}{_{\sim}}$\,1.
On the other hand $z\sim1$ is the lower limit of sub-mm/mm selected surveys \cite{sm01,d02}.
Thus samples such as ours should act as a bridge between the local Universe and the much higher redshift population detected in long-wavelength surveys. We illustrated this explicitly in Fig.~4 where we overlaied a number of SCUBA-selected sources, and showed that they occupy essentially the same sub-mm/near-IR locus as our high-$z$ candidates. 
\subsection{Conclusions}
We have learned that the brightest $\sim$10$\%$ of the CIB is composed of two different types of galaxy: about 1/6 is the low redshift tail of a rapidly evolving ULIG population (similar to the higher-$z$ SCUBA sources); and the other 5/6 are mainly nearby quiescently star-forming galaxies like the Milky Way, with perhaps a few more luminous infra-red galaxies inbetween. This is somewhat contrary to early expectations for the nature of the FIRBACK galaxies (e.g. Puget et al.~1999) where some sources in the range $z\sim1-2$ were expected. This would be the appropriate range for the handful of higher-$z$ sources in our sample only if a much warmer SED is assumed. However, this is inconsistent with the redshift and temperature estimates for the majority of our sample, ruling out such models. Further progress on constraining models in detail will come from spectroscopic and morphological studies of the entire sample.  Understanding what makes up the other $\sim$\,90\,$\%$ of the CIB will need to await future far-IR missions with smaller beam-sizes, such as BLAST and {\sl Herschel}, as well as high sensitivity mid-IR facilities such as {\sl SIRTF}.
\section*{Acknowledgments}
We wish to thank the staff of the JCMT for their assistance during the observations. This research was supported by the National Research Council (NRC) and by the Natural Sciences and Engineering Research Council of Canada. HD acknowledges funding from the MIPS project, which is supported by NASA 
through the Jet Propulsion Laboratory, subcontract \#P435236.
The James Clerk Maxwell Telescope is operated by the Joint Astronomy Centre on behalf of the Particle Physics and Astronomy Research Council of the United Kingdom, the Netherlands Organization for Scientific Research, and the National Research Council of Canada. This research has made use of the NASA/IPAC Extragalactic Database (NED) which is operated by the Jet Propulsion Laboratory, California Institute of Technology, under contract with the National Aeronautics and Space Administration.

\bsp

\label{lastpage}

\end{document}